\documentclass[11pt]{article}
\usepackage{epsfig}
\graphicspath{{/}}

\begin{document}
\baselineskip 24pt
\begin{center}
{\large\bf Plasmon Amplification through Stimulated Emission at Terahertz Frequencies in Graphene} \\
\normalsize
\vspace{5em}
Farhan Rana$^{1}$, Faisal R. Ahmad$^{2}$ \\	
\vspace*{2em}
{\em $^{1}$School of Electrical and Computer Engineering, Cornell University, Ithaca, NY 14853 \\
$^{2}$Department of Physics, Cornell University, Ithaca, NY 14853} \\
\vspace{4em}
\large\bf Abstract \\
\normalsize
\end{center}

We show that plasmons in two-dimensional graphene can have net gain at terahertz frequencies. The coupling of the plasmons to interband electron-hole transitions in population inverted graphene layers can lead to plasmon amplification through the process of stimulated emission. We calculate plasmon gain for different electron-hole densities and temperatures and show that the gain values can exceed $10^{4}$ cm$^{-1}$ in the 1-10 terahertz frequency range, for electron-hole densities in the $10^{9}$-$10^{11}$ cm$^{-2}$ range, even when plasmon energy loss due to intraband scattering is considered. Plasmons are found to exhibit net gain for intraband scattering times shorter than 100 fs. Such high gain values could allow extremely compact terahertz amplifiers and oscillators that have dimensions in the 1-10 $\mu$m range.    

\newpage 

\begin{figure}[tb]
  \begin{center}
   \epsfig{file=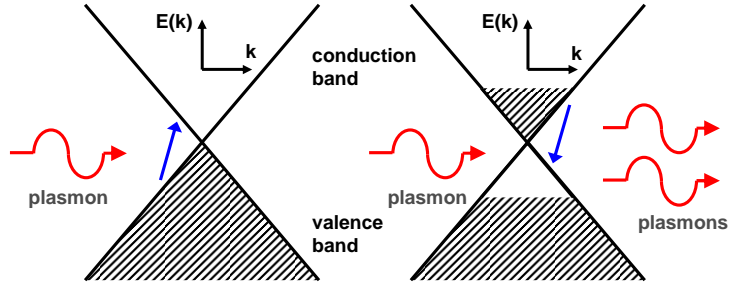,angle=-90,width=4.0 in}    
    \caption{(LEFT) Energy bands of graphene showing stimulated absorption of plasmons. (RIGHT) Population inverted graphene bands showing stimulated emission of plasmons.}
    \label{Fig1}
  \end{center}
\end{figure}

\section{Introduction}
Tremendous interest has been generated recently in the electronic properties of two dimensional (2D) graphene in both experimental and theoretical arenas~\cite{chakra, darma, ryzhii,nov1,nov2,zhang}. Graphene is a single atomic layer of carbon atoms forming a dense honeycomb crystal lattice~\cite{dressel}.  The massless energy dispersion relation of electrons and holes with zero (or close to zero) bandgap results in novel behavior of both single-particle and collective excitations~\cite{chakra,darma,ryzhii}. In addition, the high mobility of electrons in graphene has generated interest in developing novel high speed devices. Recently, it has been shown that the frequencies of plasma waves in graphene at moderate carrier densities (~$10^{9}-10^{11}$ cm$^{-2}$) are in the terahertz range~\cite{ryzhii}. Electron-hole decay through plasmon emission has been recently experimentally observed in graphene~\cite{ohta}. The zero bandgap of graphene leads to strong damping of the plasma waves (plasmons) at finite temperatures as plasmons can decay by exciting interband electron-hole pairs~\cite{chakra,darma}. In this paper we show that plasmon amplification through stimulated emission is possible in population inverted graphene layers. This process is depicted in Fig.1. We show that plasmons in graphene can have a net gain at frequencies in the 1-10 THz range even if plasmon losses from electron and hole intraband scattering are considered. A net gain for the plasmons implies that terahertz amplifiers and oscillators based on plasmon amplification through stimulated emission are possible. The gain at terahertz frequencies is possible due to the (almost) zero bandgap of graphene. Although terahertz gain is also achievable in population inverted subbands in 2D quantum wells~\cite{hu}, intrasubband plasmons in quantum wells, being longitudinal collective modes, do not couple with intersubband transitions that require field polarization perpendicular to the plane of the quantum wells. The electromagnetic energy in the two-dimensional plasmon mode is confined within very small distances of the graphene layer and therefore waveguiding structures with large dimensions, such as those required in terahertz quantum cascade lasers~\cite{hu}, are not required for realizing plasmon based terahertz devices. We also present results for plasmon gain under different population inversion conditions taking into account both intraband and interband electronic transitions and carrier scattering.  

\section{Theoretical Model}
In this section we discuss the theoretical model used to obtain the values for the plasmon gain in graphene. In graphene, the valence and conduction bands resulting from the mixing of the $p_{z}$-orbitals are degenerate at the inequivalent $K$ and $K'$ points of the Brillouin zone~\cite{dressel}. Near these points, the conduction and valence band dispersion relations can be written compactly as~\cite{darma},
\begin{equation}
E_{s,{\bf k}} = s \hbar v |{\bf k}|
\end{equation}
where $s=\pm 1$ stand for conduction ($+1$) and valence ($-1$) bands, respectively, and $v$ is the ``light'' velocity of the massless electrons and holes. The wavevector ${\bf k}$ is measured from the $K$($K'$) point. The frequencies $\omega({\bf q})$ of the longitudinal plasmon modes of wavevector ${\bf q}$ are given by the equation,$\epsilon({\bf q},\omega) = 0$, where $\epsilon({\bf q},\omega)$ is the longitudinal dielectric function of graphene~\cite{darma}. In the random phase approximation (RPA) $\epsilon({\bf q},\omega)$ can be written as~\cite{huag},
\begin{equation}
\epsilon({\bf q},\omega) = 1 - V({\bf q}) \Pi({\bf q},\omega) \label{rpa}
\end{equation}
Here, $V({\bf q})$ is the bare 2D Coulomb interaction and equals $e^{2}/2\epsilon_{\infty}q$. $\epsilon_{\infty}$ is the average of the dielectric constant of the media on either side of the graphene layer. $\Pi({\bf q},\omega)$ is the electron-hole propagator including both intraband and interband processes and is given by the expression~\cite{chakra,darma},
\begin{equation}
  \Pi({\bf q},\omega) = 4 \sum_{s\,s'\,{\bf k}} \frac{ |<\psi_{s',{\bf k}+{\bf q}}|e^{i {\bf q}.{\bf r}}|\psi_{s,{\bf k}}>|^{2} \left[ f(E_{s,{\bf k}} - E_{fs}) - f(E_{s',{\bf k}+{\bf q}} - E_{fs'}) \right]}{ \hbar \omega + E_{s,{\bf k}} - E_{s',{\bf k}+{\bf q}} + i\eta}
\end{equation}
The factor of 4 outside in the above equation comes from the degenerate two spins and the two valleys at $K$ and $K'$. $f(E-E_{f})$ is the Fermi distribution function with Fermi energy $E_{f}$. $|\psi_{s,{\bf k}})>$ are the Bloch functions for the conduction and valence bands near the $K$($K'$) point. The occupancy of electrons in the conduction and valence bands are described by different Fermi levels to allow for nonequilibrium population inversion. The Bloch functions have the following matrix elements~\cite{dressel},
\begin{equation}
 |<\psi_{s',{\bf k}+{\bf q}}|e^{i {\bf q}.{\bf r}}|\psi_{s,{\bf k}}>|^{2} = \frac{1}{2} \, \left( 1 + s s' \frac{|{\bf k}|+|{\bf q}| \cos{(\theta)}}{|{\bf k} + {\bf q}|} \right)
\end{equation}
where $\theta$ is the angle between the vectors ${\bf k}$ and ${\bf q}$. The condition $v |{\bf q}| < \omega({\bf q})$ must be satisfied in order to avoid direct intraband absorption of plasmons. Assuming $v |{\bf q}| < \omega$, and using the symmetry between conduction and valence bands, the intraband and interband contributions to the propagator can be approximated as follows,
\begin{equation}
 \Pi_{intra}({\bf q},\omega) \approx \frac{q^{2} \, K \, T\,/\pi \hbar^{2}}{ \omega (\omega + i/\tau) - v^{2}q^{2}/2} \log{\left[ \left( e^{E_{f+}/KT} + 1 \right) \left( e^{-E_{f-}/KT} + 1 \right) \right]} \label{propintra}
\end{equation}
\begin{eqnarray}
\Pi_{inter}({\bf q},\omega)  & \approx &  \frac{q^{2}}{\hbar} \int_{0}^{\infty} \frac{d{\overline \omega}}{2 \pi} \,  \frac{\left[ f(\hbar {\overline \omega}/2 - E_{f+}) -  f(-\hbar {\overline \omega}/2 - E_{f-}) \right]} {{\overline \omega}^{2} - \omega^{2}} \nonumber \\
& & + i \frac{q^{2}}{4 \hbar \omega} \left[ f(\hbar \omega/2 - E_{f+}) -  f(-\hbar \omega/2 - E_{f-}) \right]
 \label{propinter}
\end{eqnarray} 
Here, $q=|{\bf q}|$. In Equation (\ref{propintra}), the intraband contribution to the propagator is written in the plasmon-pole approximation that satisfies the f-sum rule~\cite{huag}. This approximation is not valid for large value of the wavevector $q$ when $\omega(q) \rightarrow vq$. However, in this paper we will be concerned with small values of the wavevector for which the plasmons have net gain, and therefore the approximation used in Equation (\ref{propintra}) is adequate. Plasmon energy loss due to intraband scattering has been included with a scattering time $\tau$ in the number-conserving relaxation-time approximation which assumes that as a result of scattering the carrier distribution relaxes to the local equilibrium distribution~\cite{mermin}. The real part of the interband contribution to the propagator modifies the effective dielectric constant and leads to a significant reduction in the plasmon frequency under population inversion conditions. The imaginary part of the interband contribution to the propagator incorporates plasmon loss or gain due to stimulated interband transitions. A necessary condition for plasmon gain from stimulated interband transitions is that the splitting of the Fermi levels of the conduction and valence electrons exceed the plasmon energy, i.e. $E_{f+}-E_{f-}>\hbar \omega$. But the plasmons will gave net gain only if the plasmon gain from stimulated interband transitions exceed the plasmon loss due to intraband scattering. The real and imaginary parts of the propagator in Equations (\ref{propintra}) and (\ref{propinter}) satisfy the Kramers-Kronig relations. Equations (\ref{propintra}) and (\ref{propinter}) can be used with Equation (\ref{rpa}) to calculate the real and imaginary parts of the plasmon frequency $\omega(q)$ as a function of $q$. However, from the point of view of device design, it is more useful to assume that the frequency $\omega$ is real and the propagation vector $q(\omega)$, written as a function of $\omega$, is complex. Since the charge density wave corresponding to plasmons has the form $e^{i{\bf q}.{\bf r} - i \omega t}$, the imaginary part of the propagation vector corresponds to net gain or loss. We define the net plasmon energy gain $g(\omega)$ as $-2 {\rm Imag}\{q(\omega)\}$.

\section{Results and Discussion}
In simulations we use $v=10^{8}$ cm/s and $\epsilon_{\infty}=4.0 \epsilon_{o}$ (assuming silicon-dioxide on both sides of the graphene layer)~\cite{chakra}. We assume a nonequilibrium situation, as in a semiconductor interband laser~\cite{coldren}, in which the electron and hole densities are equal and $E_{f+} = - E_{f-}$. Such a non-equilibrium situation can be realized experimentally by either carrier injection in an electrostatically defined graphene pn-junction or through optical pumping~\cite{marcus,kim}. The value of the scattering time $\tau$ (momentum relaxation time) is also critical for calculations of the net plasmon gain. Value of $\tau$ can be estimated from the experimentally reported values of mobility using the following expression for the graphene conductivity (assuming that only electrons are present)~\cite{darmatr},
\begin{equation}
\sigma =  \frac{e^{2} \,\tau \, K \, T }{\pi \hbar^{2}} \log \left( e^{E_{f+}/KT} + 1 \right) \label{condintra}
\end{equation}
Values of mobility between 20,000 and 60,000 cm$^{2}$/V-s have been experimentally measured at low temperatures (T<77K) in graphene~\cite{nov2,zhang,heer}. Assuming a mobility value of 27,000 cm$^{2}$/V-s , reported in Ref.~\cite{heer} for an electron density of $3.4\times10^{12}$ cm$^{-2}$ at T=58K, the value of $\tau$ comes out to be approximately 0.6 ps. The phonon scattering time was experimentally determined to be close to 4 ps at T=300K~\cite{heer}. Therefore, impurity or defect scattering is expected to be the dominant momentum relaxation mechanism in graphene, and the scattering time is expected to be relatively independent of temperature~\cite{darmatr}. In the results presented below, unless stated otherwise, we have used a temperature independent scattering time of 0.5 ps. 

\begin{figure}[tbp]
  \begin{center}
   \epsfig{file=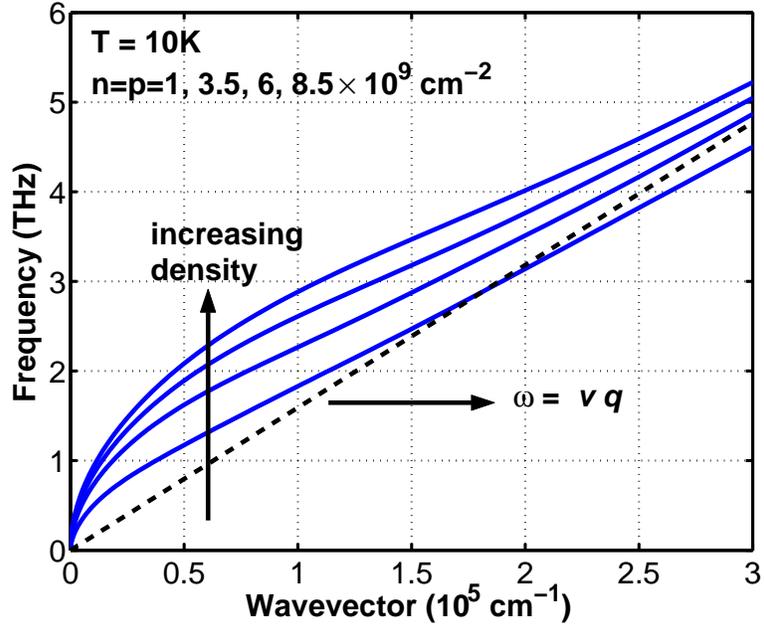,angle=0,width=4.0 in}    
    \caption{Calculated plasmon dispersion relation in graphene at 10K is plotted for different electron-hole densities ($n=p=1, 3.5, 6, 8.5 \times 10^{9}$ cm$^{-2}$). The condition $\omega({\bf q}) > \hbar v q$ is satisfied for frequencies that have net gain in the terahertz range. The assumed values of $v$ and $\tau$ are $10^{8}$ cm/s and 0.5 ps, respectively.}
    \label{Fig2}
  \end{center}
\end{figure}

\begin{figure}[tbp]
  \begin{center}
   \epsfig{file=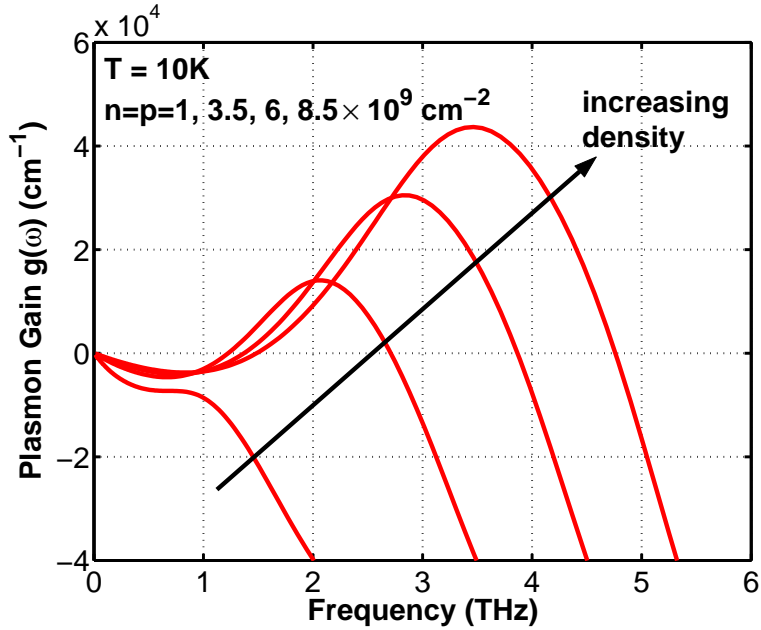,angle=0,width=4.0 in}    
    \caption{Net plasmon gain in graphene at 10K is plotted for different electron-hole densities ($n=p=1, 3.5, 6, 8.5 \times 10^{9}$ cm$^{-2}$). The assumed values of $v$ and $\tau$ are $10^{8}$ cm/s and 0.5 ps, respectively.}  
    \label{Fig3}
  \end{center}
\end{figure}

\begin{figure}[tbp]
  \begin{center}
   \epsfig{file=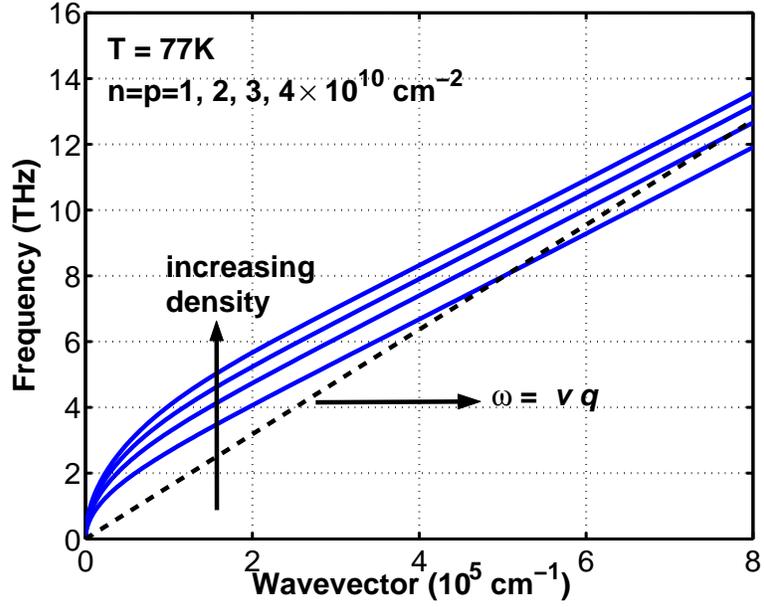,angle=0,width=4.0 in}    
    \caption{Calculated plasmon dispersion relation in graphene at 77K is plotted for different electron-hole densities ($n=p=1, 2, 3, 4 \times 10^{10}$ cm$^{-2}$). The condition $\omega({\bf q}) > \hbar v q$ is satisfied for frequencies that have net gain in the terahertz range. The assumed values of $v$ and $\tau$ are $10^{8}$ cm/s and 0.5 ps, respectively.}
    \label{Fig4}
  \end{center}
\end{figure}

\begin{figure}[tbp]
  \begin{center}
   \epsfig{file=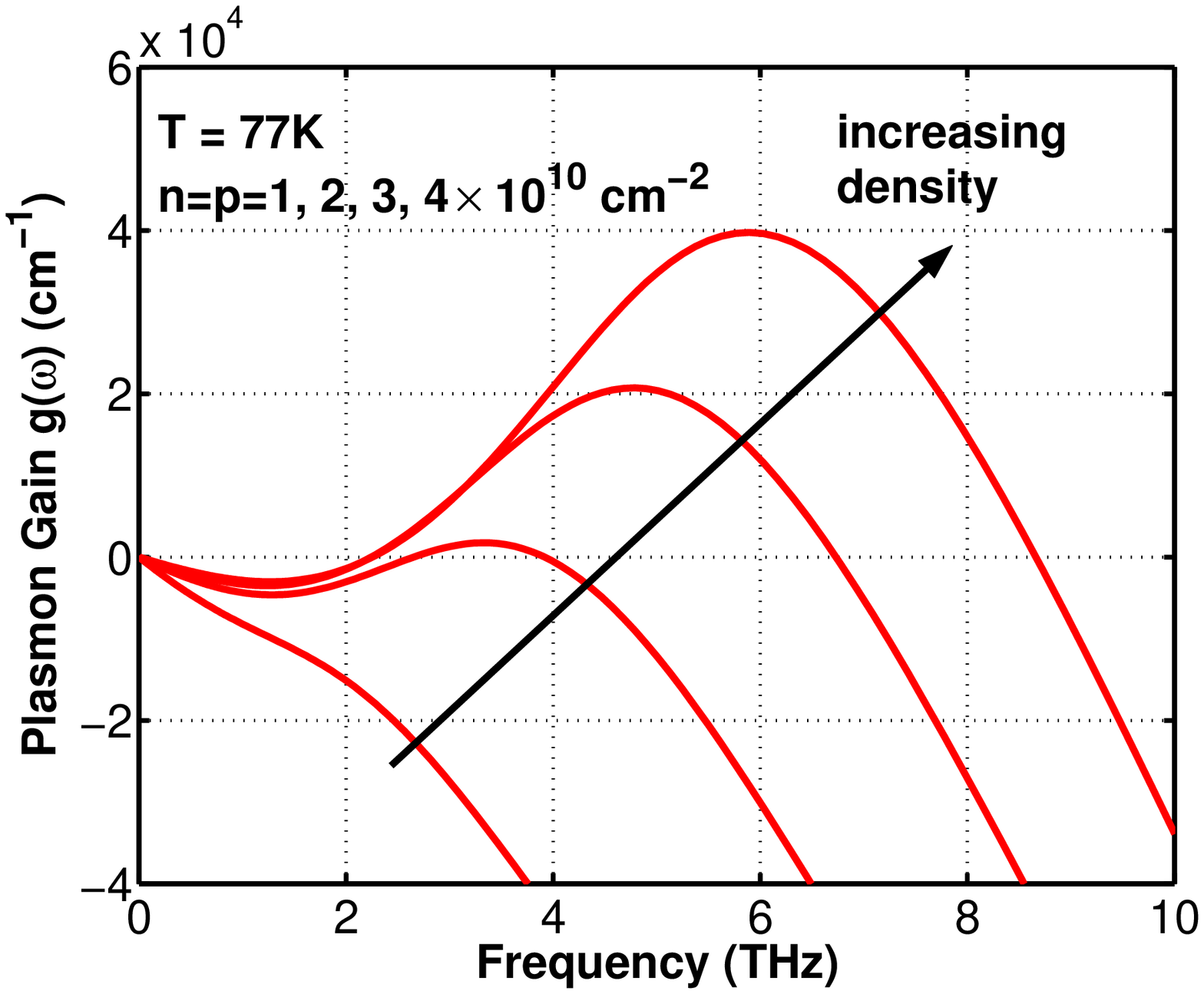,angle=0,width=4.0 in}    
    \caption{Net plasmon gain in graphene at 77K is plotted for different electron-hole densities ($n=p=1, 2, 3, 4 \times 10^{10}$ cm$^{-2}$). The assumed values of $v$ and $\tau$ are $10^{8}$ cm/s and 0.5 ps, respectively.}  
    \label{Fig5}
  \end{center}
\end{figure}

\begin{figure}[tbp]
  \begin{center}
   \epsfig{file=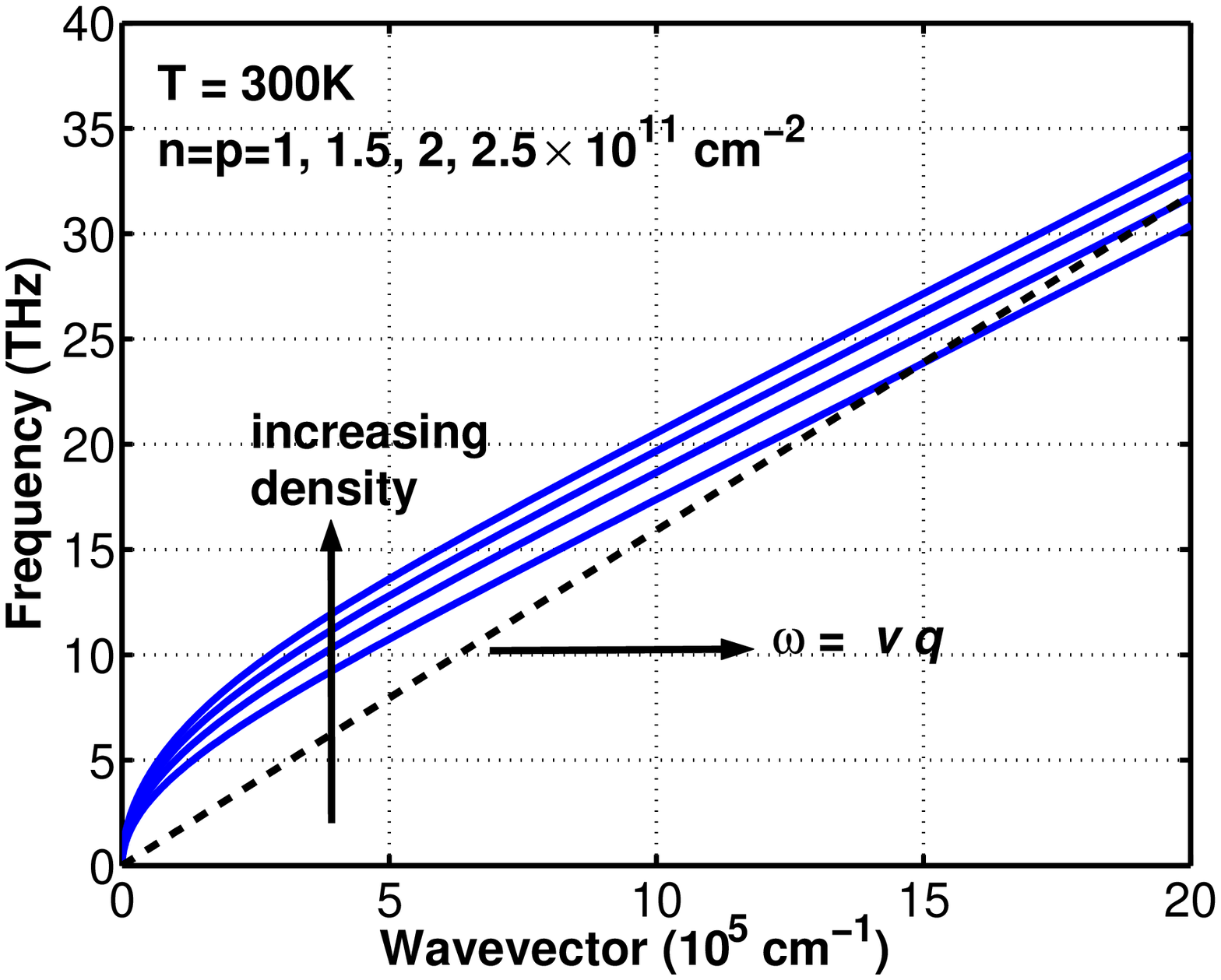,angle=0,width=4.0 in}    
    \caption{Calculated plasmon dispersion relation in graphene at 300K is plotted for different electron-hole densities ($n=p=1, 1.5, 2, 2.5 \times 10^{11}$ cm$^{-2}$). The condition $\omega({\bf q}) > \hbar v q$ is satisfied for frequencies that have net gain in the terahertz range. The assumed values of $v$ and $\tau$ are $10^{8}$ cm/s and 0.5 ps, respectively.}
    \label{Fig6}
  \end{center}
\end{figure}

\begin{figure}[tbp]
  \begin{center}
   \epsfig{file=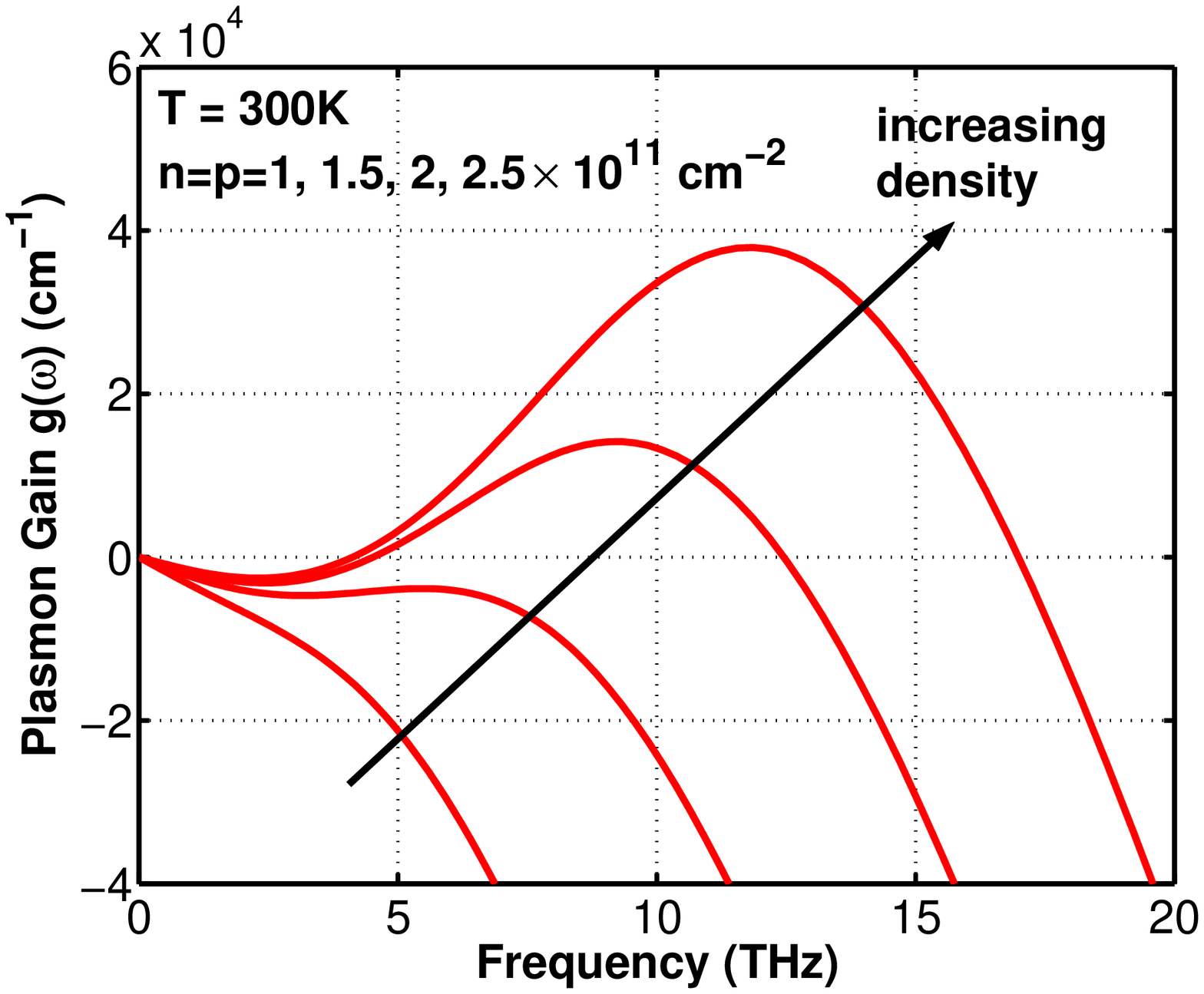,angle=0,width=4.0 in}    
    \caption{Net plasmon gain in graphene at 300K is plotted for different electron-hole densities ($n=p=1, 1.5, 2, 2.5 \times 10^{11}$ cm$^{-2}$). The assumed values of $v$ and $\tau$ are $10^{8}$ cm/s and 0.5 ps, respectively.}  
    \label{Fig7}
  \end{center}
\end{figure}

Figs.~\ref{Fig2}-\ref{Fig7} show the calculated dispersion relation of the plasmons and the net plasmon gain at T=10K, 77K, and 300K for different electron-hole densities. At very low frequencies the losses from intraband scattering dominate. At frequencies ranging from 1 to 15 THz, the plasmons can have net gain. The values of the net gain are found to be significantly large reaching $1-4 \times 10^{4}$ cm$^{-1}$ for electron-hole densities in the $10^{9}$ cm$^{-2}$ range at low temperatures and $10^{11}$ cm$^{-2}$ range at room temperature. The calculated plasmon dispersions indicate that $\omega({\bf q}) > v q $ at all frequencies for which the plasmons have net gain. Therefore, direct intraband absorption of plasmons is not possible at these frequencies and will not reduce the calculated gain values. Plasmons acquire net gain for smaller electron-hole densities at lower temperatures. At higher temperatures the distribution of electrons and holes in energy is broader and the gain at any particular frequency is therefore smaller. At T=10K, the plasmons have net gain for electron-hole densities as small as  $2 \times 10^{9}$ cm$^{-2}$. Almost an order of magnitude larger electron-hole densities are required to achieve the same net gain values at T=77K compared to T=10K. The linear energy dependence of the density of states associated with the massless dispersion relation of electrons and holes in graphene results in the maximum plasmon gain values to increase with the electron-hole density. The peak gain values shift to higher frequencies with the increase in the electron-hole density for the same reason. 


\begin{figure}[tbp]
  \begin{center}
   \epsfig{file=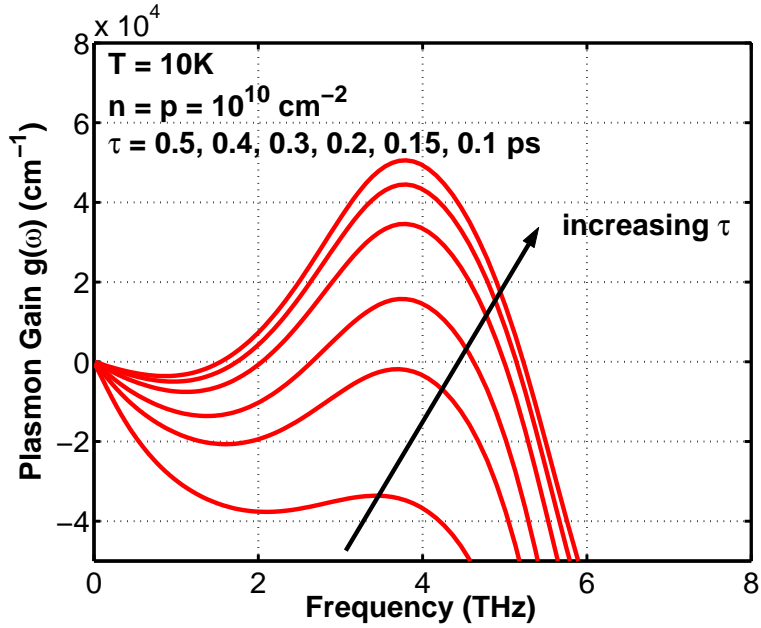,angle=0,width=4.0 in}    
    \caption{Net plasmon gain in graphene at 10K is plotted for different intraband scattering times $\tau$ ($\tau=0.5,0.4,0.3,0.2,0.15,0.1$ ps). The assumed value of $v$ is $10^{8}$ cm/s and the electron-hole density is $10^{10}$ cm$^{-2}$.}
    \label{Fig8}
  \end{center}
\end{figure}
\begin{figure}[tbp]
  \begin{center}
   \epsfig{file=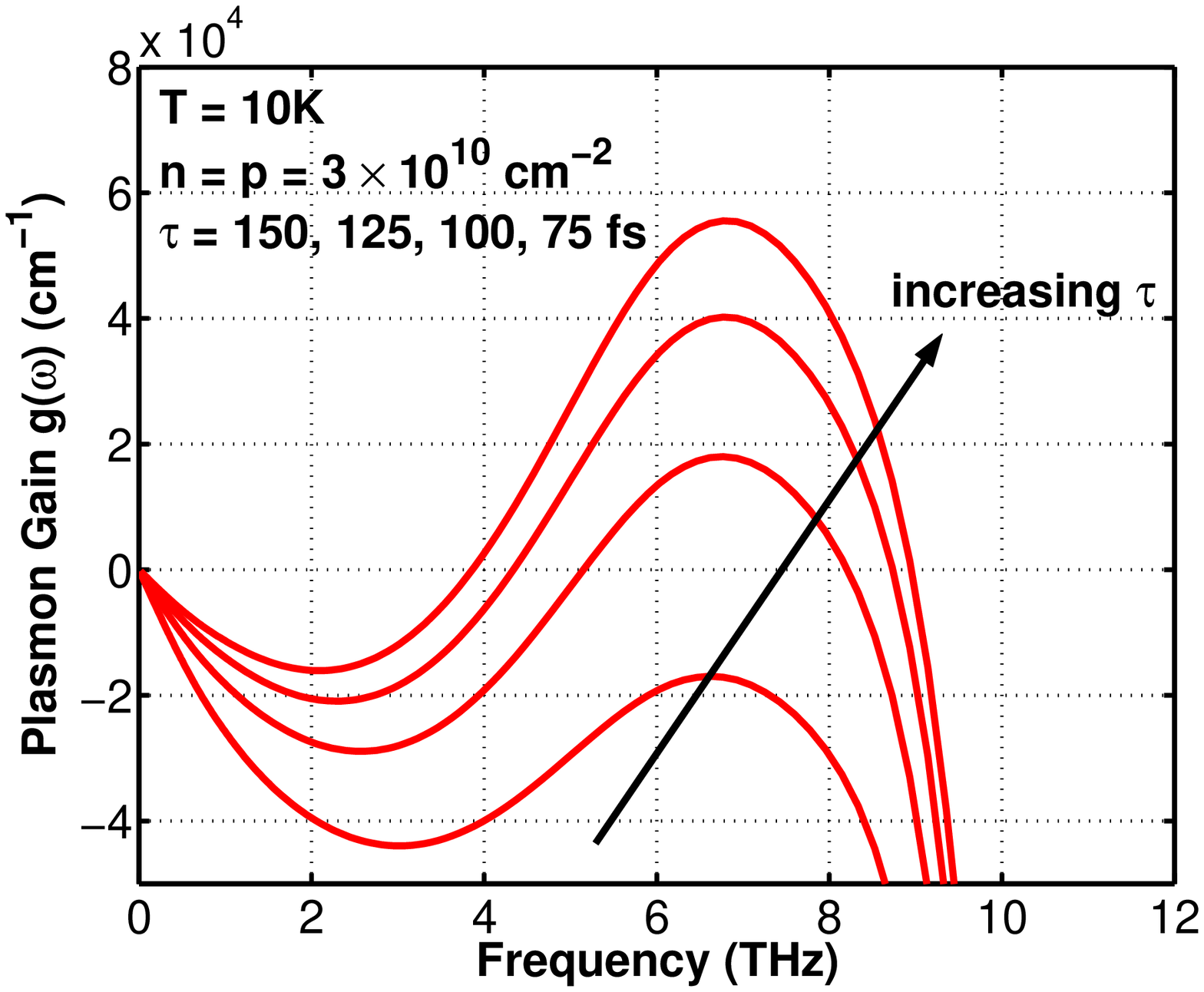,angle=0,width=4.0 in}    
    \caption{Net plasmon gain in graphene at 10K is plotted for different scattering times $\tau$ ($\tau=150,125,100,75$ fs). The assumed value of $v$ is $10^{8}$ cm/s and the electron-hole density is $3\times 10^{10}$ cm$^{-2}$.}
    \label{Fig9}
  \end{center}
\end{figure}

The fact that plasmons can acquire net gain for relatively small carrier densities suggests that plasmon gain is relatively robust with respect to intraband scattering losses. Fig.~\ref{Fig8} shows the net gain at T=10K for $n=p=10^{10}$ cm$^{-2}$ and values of the intraband scattering time $\tau$ varying from 0.1 to 0.5 ps. The net gain decreases as the plasmon losses increase with a decrease in the value of $\tau$ and the maximum gain value equals zero for $\tau=0.15$ ps. However, it should not be concluded from Fig.~\ref{Fig8} that plasmons cannot have net gain for $\tau$ less than 0.15 ps since electron-hole density can always be increased to achieve net gain for smaller values of $\tau$.  Fig.~\ref{Fig9} shows the net gain at T=10K for $n=p=3\times10^{10}$ cm$^{-2}$ and values of the intraband scattering time $\tau$ varying from 75 to 150 fs. It can be seen that at these larger carrier densities plasmons have net gain for scattering times that are sub-100 fs. 

The exceedingly large values of the net plasmon gain ($>10^{4}$ cm$^{-1}$) in graphene implies that terahertz plasmon oscillators only a few microns long in length could have sufficient gain to overcome both intrinsic losses and losses associated with external radiation coupling. Plasmon fields with in-plane wavevector magnitude $q$ decay as $e^{-q\,|z|}$ away from the graphene layer where $|z|$ is the distance from the graphene layer. Figs.~\ref{Fig2},~\ref{Fig4}, and ~\ref{Fig6} show that $q$ has values exceeding $10^{5}$ cm$^{-1}$ at terahertz frequencies. Therefore, the electromagnetic energy associated with the terahertz plasmons is confined within 100 nm of the graphene layer. Strong field confinement and low plasmon losses at terahertz frequencies are both partly responsible for the high net gain values in graphene. Recent theoretical predictions for electron-hole recombination rates in graphene due to Auger scattering indicate that electron-hole recombination times can be much longer than 1 ps at temperatures ranging from 10K to 300K for electron-hole densities smaller than $10^{12}$ cm$^{-2}$~\cite{rana}. This suggests that population inversion can be experimentally achieved in graphene via current injection in electrostatically defined pn-junctions or via optical pumping~\cite{marcus,kim}. It also needs to be pointed out here that graphene monolayers and multilayers produced from currently available experimental techniques are estimated to have defect/impurity densities anywhere between $10^{11}$ and $10^{12}$ cm$^{-2}$~\cite{darmatr}. Therefore, at low electron-hole densities (less than $10^{11}$ cm$^{-2}$) graphene is expected to exhibit localized electron and hole puddles rather than continuous electron and/or hole sheet charge densities~\cite{darmatr}. This implies that with the currently available techniques graphene based terahertz plasmon oscillators might only be realizable with higher electron-hole densities ($>10^{11}$ cm$^{-2}$) for operation at higher frequencies ($>5$ THz). 

\section{conclusion}
In conclusion, we have shown that high gain values for plasmons are possible in population inverted graphene layers in the 1-10 THz frequency range. The plasmon gain remains positive even for carrier intraband scattering times shorter than 100 fs. The high gain values and the strong plasmon field confinement near the graphene layer could enable compact terahertz amplifiers and oscillators. The authors would like to thank Edwin Kan and Sandip Tiwari for helpful discussions.

\end{document}